# Dissipative Transport in Rough Edge Graphene Nanoribbon Tunnel Transistors

Youngki Yoon[†] and Sayeef Salahuddin[†]

Department of Electrical Engineering and Computer Sciences, University of California, Berkeley, CA 94720, USA

**ABSTRACT**

We have studied quantum transport in Graphene Nanoribbon Tunnel Field-Effect Transistors. Unlike other studies on similar structures, we have included dissipative processes induced by inelastic electron-phonon scattering and edge roughness in the nanoribbon self-consistently within a non-equilibrium transport simulation. Our results show that the dissipative scattering imposes a limit to the minimum OFF current and a minimum subthreshold swing that can be obtained even for long channel lengths where direct source-drain tunneling is inhibited. The edge roughness, in presence of dissipative scattering, somewhat surprisingly, shows a classical behavior where it mostly reduces the maximum ON current achievable in this structure.

[†]E-mail: yyoon@eecs.berkeley.edu, sayeef@eecs.berkeley.edu

Tunnel Field Effect Transistors (TFETs)[1,2] are currently being considered as a potential candidate to replace Si CMOS for low power applications. Graphene Nanoribbon (GNR) has been identified as one of the promising material systems for a TFET due to the planar structure, potential for electrostatic doping, amenability to strong electrostatic control and 1-D nature of transport. Indeed, very abrupt switching behavior of GNR TFETs has been predicted.[3,4] Recent progress in obtaining highly smooth edge GNRs[5,6] also suggests that GNR based TFET structures may ultimately be feasible.

Due to the fact that quantum-mechanical band-to-band tunneling is the dominant transport mechanism in a TFET, it is of critical importance to understand the underlying physics of tunneling as a function of electric field, device geometry and doping conditions. Previously, a number of papers have investigated these problems by approximate, semi-analytic and more rigorous, numerical Non-Equilibrium Green's Function (NEGF) methods.[3,4,7] Most of these studies, however, were done within an approximation of ballistic transport that ignores dissipative processes and their influence on the current-voltage characteristic. An improvement over these ballistic calculations was done by Lu *et al.* who studied the role of electron-phonon coupling and associated inelastic scattering within a mode space approach.[8] A significant difference was found in device behavior compared to the ballistic picture. In nature, this study is similar to previous phonon-limited transport studies for carbon nanotube based tunnel FETs.[9] However, a mode-space approach, while numerically cost-effective, cannot account for the inter-band scattering events. A real-space approach, which can treat such inter-band scattering in a rigorous manner, has recently been reported by these authors for GNR MOSFETs.[10]

However, transport in GNR is not only affected by electron-phonon scattering but also influenced by edge roughness. A recent simulation study, that includes statistical edge roughness but not phonon scattering, indicated that edge roughness can increase the OFF current of a GNR TFET by a few orders of magnitude.[11] This is very different from surface roughness scattering in conventional transistors where the main role of such scattering is to reduce the drive current of the MOSFET. Then a relevant question would be, what is the true nature of the transport current when both electron-phonon scattering and edge roughness are present? To answer this, one needs to solve the NEGF equation self-consistently with phonon scattering and random potential due to edge roughness, repeat many times with samples having different edge shapes, and then take an average so that a statistical trend for edge roughness emerges. While the theoretical formalism to do such a study has been established, the tremendous numerical burden makes it an extremely difficult problem to undertake.

In this paper, we perform such self-consistent NEGF simulations by including electron-phonon scattering in the real space, thus accounting for both intra-band and inter-band events simultaneously, and also edge roughness of the ribbons. Our results show, somewhat surprisingly, that electron-phonon scattering dominates the qualitative device behavior and the main contribution of edge roughness is very much classical, i.e., to reduce the drive current of the device. Our results also show that phonon scattering imposes a minimum limit of the OFF current and thereby the minimum subthreshold swing that can be obtained in a given tunnel FET structure.

The schematic cross-section of the simulated device is shown in Fig. 1(a). $N$=16 armchair-edge GNR is used for an active channel material. Uniform width with ideal edge is assumed first to assess the roughness-free performance and scaling trends, and then non-ideal edge is introduced to study the performance variation. The device structure is like a simple lateral tunnel transistor where source is p-doped and drain is n-doped.[12,13] For gate insulator, an effective oxide thickness (EOT) of ~1 nm (2.5 nm $Al_2O_3$) gate insulator is assumed. All calculations also assume room temperature (300 K).

To describe the carrier transport through GNR TFETs, we solve the Schrödinger equation by the NEGF formalism with tight-binding approximation in a $p_z$ orbital basis set. Phonon scattering is treated with the self-consistent Born approximation by calculating electron-phonon coupling constant from the perturbation potential.[10,14,15] We only treat on-site scattering.[16] The transport equation is solved iteratively with the Poisson's equation for the self-consistency between charge density and electrostatic potential.

In preparing the samples for rough-edge simulations, it is assumed that the edge of GNR is smooth enough to allow to model it in such a way that one carbon atom out of hundred sites on the edges is either missing or added (1% of the roughness). For example, for a 70-nm GNR (the length used in this study), the roughness appears every 7 Å in average on both sides of the edges. However, we did not simply take random positions for the roughness, but dependence and correlation between the non-idealities were taken into account in generating the samples. The reason behind this is that the roughness is locally correlated in practice when GNR is obtained by unzipping a CNT[17,18] or by etching or patterning a graphene. Figure 1(b) shows a sample of GNR with roughness where one may see six consecutive missing atoms (strong correlation) or two consecutive additional atoms (weak correlation) at different positions. Here we used 50% of probability for the same disorder to happen if the adjacent carbon atom is missing or added.

First of all, we investigate how switching characteristic is affected by phonons in GNR TFETs. For this, we performed two separate simulations, in one of which ballistic transport is assumed [Fig. 2(a)] and phonon scattering is considered in the other [Fig. 2(b)], by varying the channel length from 10 to 30 nm. The main difference is that, in case of ballistic transport, minimum leakage current ($I_{min}$) exponentially decreases with increasing channel length, while that in dissipative transport is almost invariant if the channel length is larger than 20 nm. This can be explained by local density-of-states (LDOS) and energy-resolved current plots in Fig. 3. In the ballistic case, the leakage current is the consequence of direct source-drain tunneling through the entire channel [Figs. 3(a) and 3(d)]. Therefore, $I_{min}$ is strongly correlated to the physical channel length. On the other hand, in the presence of phonon scattering, there exists anther leakage path, which results from the gap states induced by optical phonons, as illustrated in Fig. 3(b). When the channel length is long enough ($L_{ch} \geq 20$ nm) where direct source-to-drain tunneling current is minimal, majority of the leakage is attributed to the phonon-assisted band-to-band tunneling [Figs. 3(b) and 3(e)]. Note that this leakage is independent to the physical channel length and hence there is effectively no difference in the subthreshold current [for the devices with $L_{ch} \geq 20$ nm in Fig. 2(b)]. However, if the channel length is short ($L_{ch} \leq 15$ nm), leakage current significantly increases in both cases by decreasing channel length. This is because the direct source-to-drain tunneling current is now dominant over the phonon-assisted band-to-band tunneling current [Figs. 3(c) and 3(f)].

What we can further infer from Fig. 2 is that, due to the phonon scattering, there exists a limit of how steep the subthreshold swing can be. Figure 4 shows minimum leakage current (in the inset) and subthreshold swing [$SS = \partial V_G/\partial(\log_{10} I_D)$] with the channel length varying from 10 to 100 nm. As discussed earlier, the minimum leakage current is monotonically reduced initially as channel length increases but soon becomes flat ($I_{min}$ = 0.37 pA) where phonon-assisted band-to-band tunneling is dominant with negligible direct source-drain leakage posed by a huge tunnel barrier (red curve), which is totally different to what is predicted by the ballistic picture (blue dashed line). Subthreshold swing also shows the same trend with respect to the channel length variation (the minimum $SS$ = 34 mV/dec). This indicates, importantly, that the minimum leakage current and subthreshold swing of GNR TFETs are limited by phonon scattering rather than channel length or density-of-state switching. Before moving on to rough-edge simulations, let us summarize the findings for ideal-edge GNR TFETs. When the channel length is less than ~20 nm, the direct source-drain tunneling decreases the OFF current exponentially with increasing channel length such that the switching behavior is significantly improved. However, this cannot be applied to the transistors larger than 20 nm to achieve extremely small OFF current. Beyond that gate length, inelastic phonon scattering takes over and essentially fixes the OFF current to a certain value. Thus ideal, highly abrupt, subthreshold swing observed in a purely ballistic calculation is washed away by phonon scattering and much smaller $I_{on}/I_{off}$ ratio is expected for the same supply voltage window.

Now, to take into account edge roughness, ten samples having different edge shapes are prepared following the method described above. One key question that we want to address here is whether a GNR TFET could still work as a switching device even with variations imposed by edge roughness. Therefore, we focus on two bias points at $V_G$ = 0.15 and 0.45 V to examine how ON and OFF characteristics and ON/OFF current ratio are affected. The circles in Fig. 5 show the deviation of device characteristics from the ideal cases due to the roughness. Our simulation results indicate that the role of phonon is critical at OFF state, while the roughness does not add significant performance degradation [Fig. 5(a)]. Although the-worst-case $I_{off}$ is larger than the ideal case by more than one order of magnitude, for the most cases, the values are not significantly affected. In fact, the average $I_{off}$ of the ten samples [shown by the lateral bar at the inset in Fig. 5(a)] is only 2.5 times larger as compared to the case without edge roughness. This result can be explained by looking into the LDOS of the device. As discussed earlier [in Fig. 3(b)], the main leakage mechanism is phonon-assisted band-to-band tunneling for a device with a reasonably long channel length. Therefore, the leakage current will be increased if the localized states caused by the roughness are created in such a way that the phonon-assisted band-to-band tunneling is enhanced. Figure 6(a) is the case that we observed the largest $I_{off}$ among the ten samples where the states at the channel-drain junction (marked with an oval) make phonon-assisted band-to-band tunneling more significant. However, note that the states in the middle of the channel do not significantly increase the leakage current since the direct source-to-drain tunneling current is much smaller than phonon-assisted tunneling current and hence the current increase by these states is still negligible. It must be pointed out that the atomistic configuration of roughness at two different positions in Fig. 6(a) is exactly the same. Therefore, it is clear that even the same roughness does not affect device performance in a similar way, but the specific position where roughness appears can be more critical.

By contrast, at the ON state, the same degree of edge roughness can significantly affect $I_{on}$ [Fig. 5(b)]. Note that, just like what happens in a conventional GNR FET, optical phonon has only negligible impacts on the DC current at the relevant voltage since electrons that emit phonons inside the drain do not have sufficient energy to travel back to the channel.[10] In this case, the effect of acoustic phonon is also small due to the relatively short channel (30 nm), which of course will become more significant as the channel length increases.[10] On the other hand, edge roughness has a large impact on the variation of $I_{on}$. According to our simulations with ten samples, the lowest value is roughly one order of magnitude smaller than the ideal case, and it shows 47% reduction of $I_{on}$ in average [the mean value of the ten is shown by the lateral bar at the inset of Fig. 5(b)]. While the mechanism to cause the performance degradation at OFF state is enhanced phonon-assisted band-to-band tunneling induced by localized states, perturbation of the potential is the key factor to limit the $I_{on}$. Figure 6(b) is the energy-resolved current spectrum inside the device for the lowest value of $I_{on}$, which shows that even small potential perturbation in the channel region (indicated by an arrow) can significantly reduce the $I_{on}$ by imposing additional barrier. However, the potential fluctuation inside the source or the drain has much less effect as is observed in Fig. 6(c) where $I_{on}$ is 90% of the ideal value. It must be noted that there is no clear trend of the values of $I_{on}$ and $I_{off}$ on the edge roughness: While the sample used in Fig. 6(c) results in the largest $I_{on}$ among the ten samples, its $I_{off}$ (1.11 pA) is close to the mean value.

To summarize, we have performed atomistic, non-equilibrium quantum transport simulations for GNR TFETs accounting for dissipative processes and edge roughness self-consistently. When the gate length is less than ~20 nm, the direct source-drain tunneling severely limits the switching characteristics. On the other hand, when the channel length is increased, while direct source-to-drain tunneling is effectively inhibited, phonon scattering sets the minimum OFF current and thereby the minimum subthreshold swing that can be obtained. In the presence of irregular edges in GNR, the effect of roughness is largely classical, where it mostly reduces the ON current of the device.

## ACKNOWLEDGMENT

This work was supported in part by FCRP center on Functional Engineered and Nano Architectonics (FENA). Computing resources were provided by DOE National Energy Research Scientific Computing Center (NERSC) under the NISE program and also by NSF Network for Computing Nanotechnology (NCN).

**FIGURES and CAPTIONS**

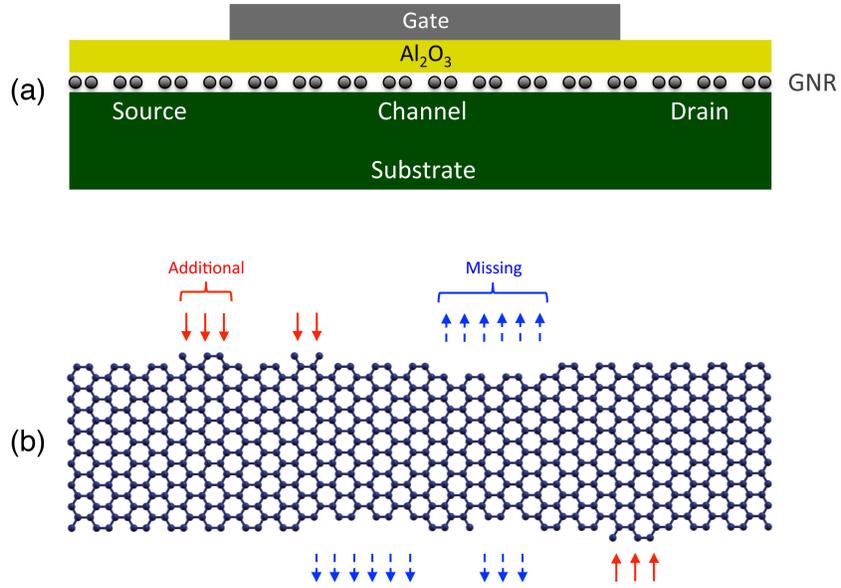

**Fig. 1.** (a) Graphene nanoribbon (GNR) tunnel field-effect transistor (TFET). $N = 16$ (width ~2 nm) armchair-edge GNR ($E_g = 0.7$ eV) is used for the active material. Source (drain) is p-doped (n-doped) by 0.5 dopant per nm (or equivalently $2.6 \times 10^{13}$ /cm$^2$). Channel length is varied from 10 to 100 nm. Source (drain) extension is 20 nm. Gate dielectric is 2.5 nm Al$_2$O$_3$ ($\kappa = 9.1$). Power supply voltage ($V_{DD}$) is 0.3 V. (b) A sample of rough edge GNR. Red solid and blue dashed arrows point out the positions where carbon atoms are added to or missing from the edges of GNR, respectively.

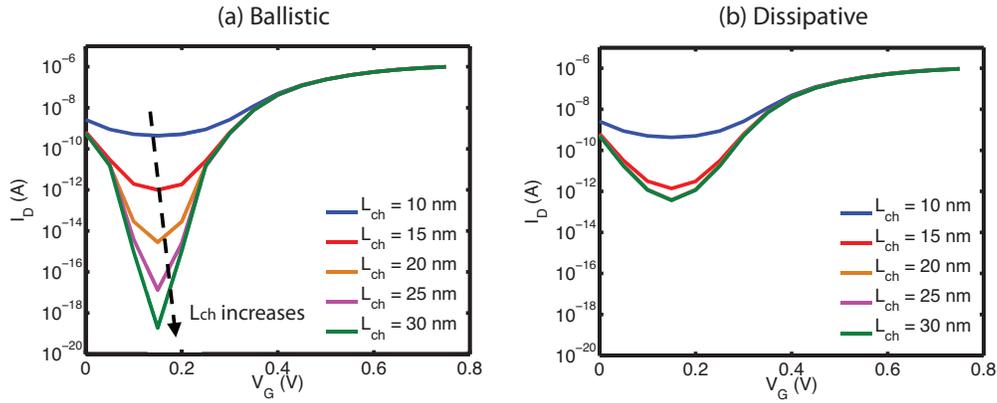

**Fig. 2.** $I_D - V_G$ plots for (a) ballistic and (b) dissipative transport. While, in case of ballistic transport, minimum leakage current exponentially decreases with increasing channel length, that under dissipative transport remains almost same for the channel length longer than 20 nm. For short channel devices ($L_{ch} \leq 15$ nm), however, leakage current significantly increases in both cases. In (b), curves for the $L_{ch}$ of 20, 25 and 30 nm cannot be distinguished due to the overlap.

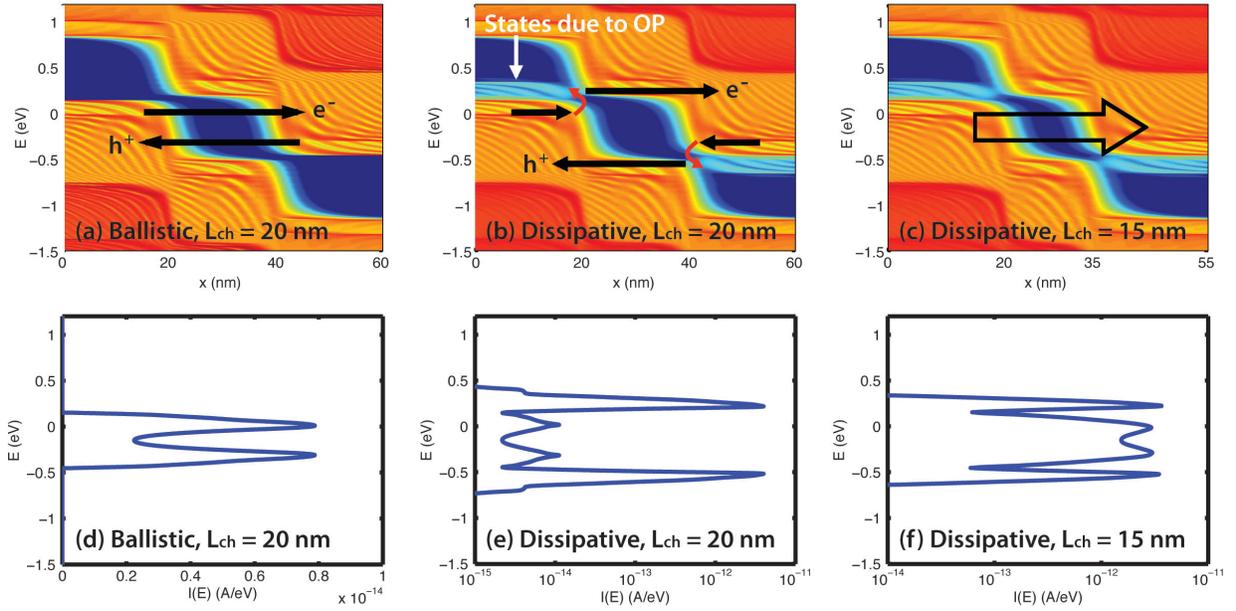

**Fig. 3.** Main leakage path shown on local density-of-states (LDOS) plot for (a) ballistic transport, dissipative transport for (b) relatively long-channel ($L_{ch}$ = 20 nm) and (c) short-channel device ($L_{ch}$ = 15 nm). The red (blue) represents more (less) density-of-states (The dark blue region clearly shows the bandgap profile across the device). In the presence of phonon scattering, phonon-assisted band-to-band tunneling is dominant over direct source-drain tunneling if the channel length is long enough. (d-f) Energy-resolved current plotted at the mid-channel for the same conditions shown in (a-c).

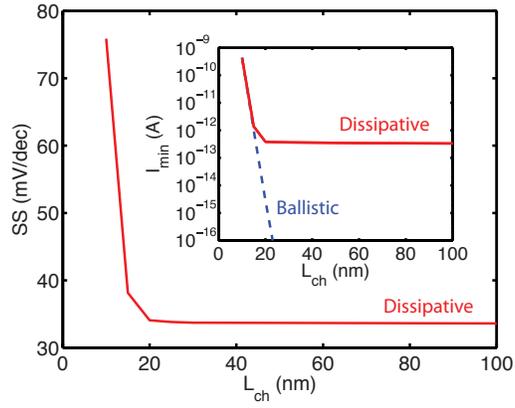

**Fig. 4.** Subthreshold swing (*SS*) and minimum leakage current ($I_{min}$) vs. channel length ($L_{ch}$). While ballistic transport simulation predicts the monotonic decrease of $I_{min}$ by increasing $L_{ch}$, dissipative transport simulation shows clear limitation of channel length modulation in reducing leakage current (inset) and subthreshold swing (main plot). Under the given conditions, $I_{min}$ is 0.37 pA and minimum subthreshold swing is 34 mV/dec.

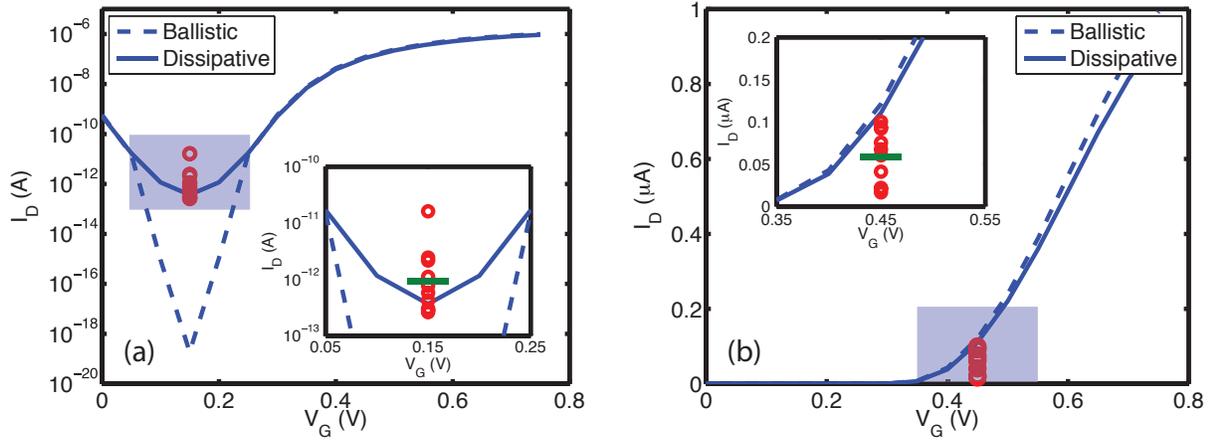

**Fig. 5.** Performance variation due to edge roughness. Device characteristics with 30-nm channel are simulated using ten GNR samples having different edge configurations (shown by circles) at the gate voltage of (a) 0.15 V and (b) 0.45 V. Ballistic (dashed line) and dissipative (solid line) simulation results with ideal edges are shown for references. The regions in the shaded boxes are enlarged at insets where lateral bars show mean values of $I_{off}$ and $I_{on}$.

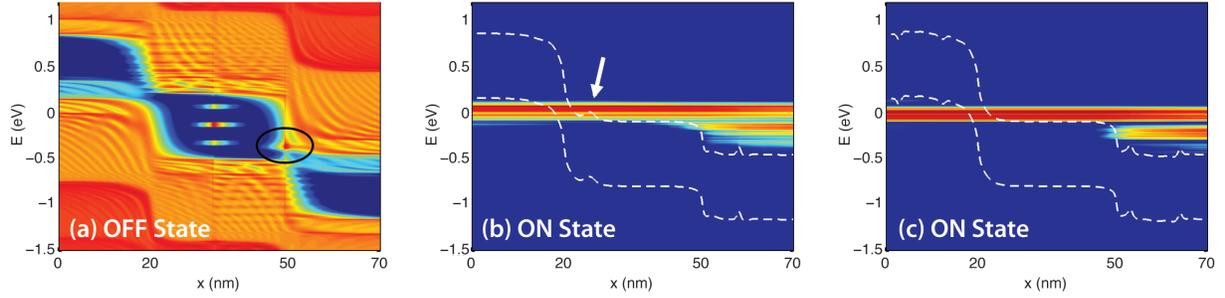

**Fig. 6.** (a) Local density-of-states for the case of the largest leakage current ($I_{off}$ = 16.2 pA) among the ten samples shown in Fig. 5(a). Localized states at the channel-drain junction (marked by an ellipse) increase leakage current significantly. (b) Energy-resolved current spectrum along the device for the smallest ON current ($I_{on}$ = 16.5 nA) among the cases shown in Fig. 5(b). Dashed lines show the conduction band and the valence band profile. The arrow indicates potential perturbation in the channel region, which reduces $I_{on}$ significantly. (c) The same is shown for the largest $I_{on}$ (0.1 µA) where potential perturbation in the source and the drain region does lightly affect the current.